\documentclass[aps,prl,twocolumn,showpacs,superscriptaddress]{revtex4-1}
\usepackage{graphicx}  
\usepackage{dcolumn}   
\usepackage{bm}        
\usepackage{amssymb}   
\usepackage{amsmath}
\usepackage{xspace}
\usepackage{xcolor}
\usepackage{atbegshi,picture}

\def\Offline{\mbox{$\overline{\textrm{Off}}$\hspace{.05em}\protect\raisebox{.4ex}{$\protect\underline{\textrm{line}}$}}\xspace}

\begin{document}

\AtBeginShipoutNext{\AtBeginShipoutUpperLeft{\put(10mm\relax,-10mm){Published in PRL as DOI:10.1103/PhysRevLett.117.192001}}}

\title{Testing Hadronic Interactions at Ultrahigh Energies \\ with Air Showers Measured by the Pierre Auger Observatory }

\author{A.~Aab}
\affiliation{Universit\"at Siegen, Fachbereich 7 Physik -- Experimentelle Teilchenphysik, Germany}

\author{P.~Abreu}
\affiliation{Laborat\'orio de Instrumenta\c{c}\~ao e F\'\i{}sica Experimental de Part\'\i{}culas -- LIP and Instituto Superior T\'ecnico -- IST, Universidade de Lisboa -- UL, Portugal}

\author{M.~Aglietta}
\affiliation{Osservatorio Astrofisico di Torino (INAF), Torino, Italy}
\affiliation{INFN, Sezione di Torino, Italy}

\author{E.J.~Ahn}
\affiliation{Fermi National Accelerator Laboratory, USA}

\author{I.~Al Samarai}
\affiliation{Laboratoire de Physique Nucl\'eaire et de Hautes Energies (LPNHE), Universit\'es Paris 6 et Paris 7, CNRS-IN2P3, France}

\author{I.F.M.~Albuquerque}
\affiliation{Universidade de S\~ao Paulo, Inst.\ de F\'\i{}sica, S\~ao Paulo, Brazil}

\author{I.~Allekotte}
\affiliation{Centro At\'omico Bariloche and Instituto Balseiro (CNEA-UNCuyo-CONICET), Argentina}

\author{J.D.~Allen}
\affiliation{New York University, USA}

\author{P.~Allison}
\affiliation{Ohio State University, USA}

\author{A.~Almela}
\affiliation{Instituto de Tecnolog\'\i{}as en Detecci\'on y Astropart\'\i{}culas (CNEA, CONICET, UNSAM), Centro At\'omico Constituyentes, Comisi\'on Nacional de Energ\'\i{}a At\'omica, Argentina}
\affiliation{Universidad Tecnol\'ogica Nacional -- Facultad Regional Buenos Aires, Argentina}

% ...

\collaboration{The Pierre Auger Collaboration}
\email{auger\_spokespersons@fnal.gov}
\homepage{\url{http://www.auger.org}}
\noaffiliation

\received{23 April 2016}

\begin{abstract}
Ultrahigh energy cosmic ray air showers probe particle physics at energies beyond the reach of accelerators.  Here we introduce a new method to test hadronic interaction models without relying on the absolute energy calibration, and apply it to events with primary energy 6-16 EeV ($E_{\rm CM}$ = 110-170 TeV), whose longitudinal development and lateral distribution were simultaneously measured by the Pierre Auger Observatory.  The average hadronic shower is $1.33 \pm 0.16$ ($1.61 \pm 0.21$) times larger than predicted using the leading LHC-tuned models EPOS-LHC (QGSJetII-04), with a corresponding excess of  muons.
\end{abstract}

\pacs{Pierre Auger Observatory, ultrahigh energy cosmic rays, muons, hadronic interactions}

\maketitle

\section{Introduction}

For many years there have been hints that the number of muons in ultrahigh energy cosmic ray (UHECR) air showers is larger than predicted by hadronic interaction models, e.g., \cite{HiRes-MIAmuons}.  Most recently, the Pierre Auger Observatory~\cite{augerNIM15}  compared the muon number in highly inclined events to predictions using the two leading  LHC-tuned hadronic event generators (HEGs) for air showers, QGSJet-II-04~\cite{QII-04,QII} and EPOS-LHC~\cite{EPOS-LHC,EPOS}.   The observed number of muons for $10^{19}$ eV primaries was found \cite{augerHorizMuons15} to be 30\%-80\% higher than the models predict assuming the primary composition inferred from the depth-of-shower-maximum distribution for each given model~\cite{augerXmaxMeas14,augerXmaxInterp14}, but the significance of the inferred muon excess is limited due to the uncertainty in the absolute energy calibration. 

For a given primary energy and mass, the number of muons is sensitive to hadronic interactions.  Typically about 25\% of the final state energy in each hadronic interaction is carried by $\pi^{0}$'s, which immediately decay to two photons and thus divert energy from the hadronic cascade, which is the main source of muons, to the electromagnetic (EM) cascade.   The hadronic cascade terminates when the energy of charged pions drops low enough that they decay before interacting, $\mathcal O(100$ GeV).  If the average fraction of EM energy per interaction were increased or decreased, or there were more or fewer generations of hadronic interactions in the cascade (which depends on the primary mass and properties of the final states such as multiplicity), the muon ground signal would be lower or higher.  Therefore, a significant discrepancy between observed and predicted muon ground signal would indicate that the description of hadronic interactions is inaccurate, assuming that the composition can be properly understood.

\begin{figure}[ht]
\centering
\includegraphics[width=\linewidth]{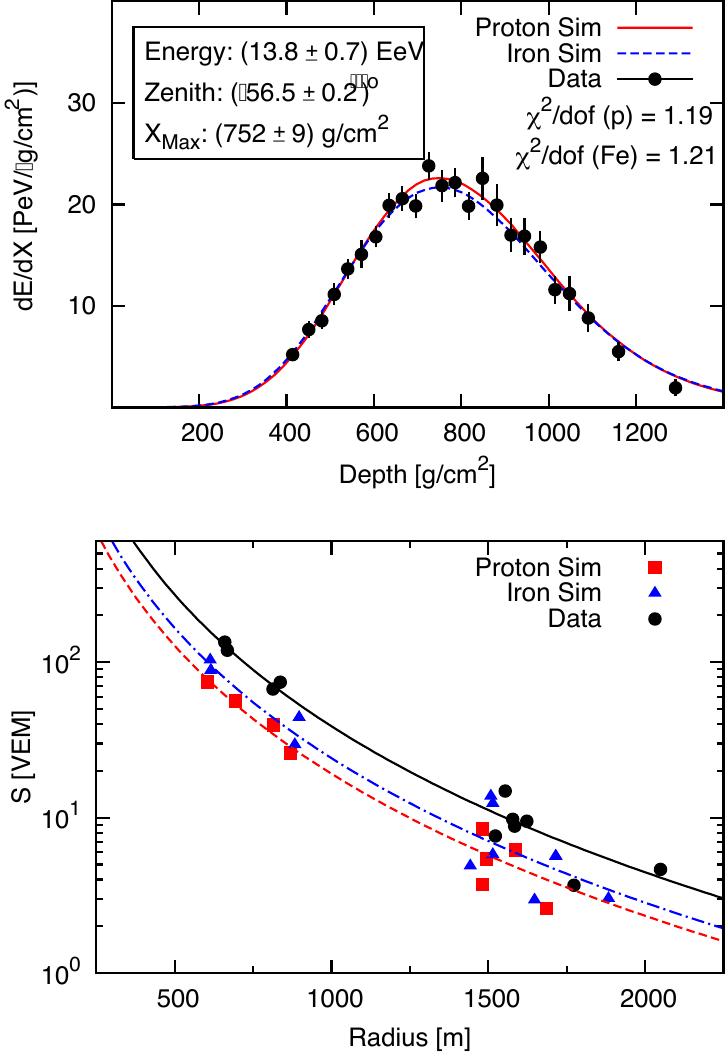}
\caption{Top: The measured longitudinal profile of an illustrative air shower with its matching simulated showers, using QGSJet-II-04 for proton (red solid) and iron (blue dashed) primaries.  Bottom: The observed and simulated ground signals for the same event (p: red squares, dashed-line, Fe: blue triangles, dot-dash line) in units of vertical equivalent muons; curves are the lateral distribution function (LDF) fit to the signal.}
\label{figFDSDComp}
\end{figure}

There has been excellent recent progress in composition determination~\cite{augerXmaxMeas14,augerXmaxInterp14,augerTAwg}, which provides a valuable ``prior" for modeling individual showers.  Here we complement that progress with a new, more powerful approach to the muon analysis which removes the sensitivity to the absolute energy calibration.  It is applicable to the entire data set of hybrid events: those events whose longitudinal profile (LP) is measured by the Pierre Auger Observatory's fluorescence detector (FD) \cite{augerFD,augerNIM15} at the same time the ground signal is measured with its surface detector (SD) \cite{augerSD,augerNIM15}.  

The ground signal of an individual shower of a CR of given energy and mass, depends primarily on the zenith angle and the depth-of-shower-maximum, $X_{\rm max}$, because together these determine the path length and thus attenuation of the electromagnetic and muonic components at ground.  
In order to most simply characterize a possible discrepancy between the predicted and observed properties of the air shower, we introduce an energy rescaling parameter, $R_{E} $, to allow for a possible shift in the FD energy calibration, and a multiplicative rescaling of the hadronic component of the shower by a factor $R_{\rm had}$.  $R_{E}$ rescales the total ground signal of the event approximately uniformly, while $R_{\rm had}$ rescales only the contribution to the ground signal of inherently hadronic origin, which consists mostly of muons.   Because the EM component of the shower is more strongly attenuated in the atmosphere than the muonic component, and the path length in the atmosphere varies as a function of zenith angle,  $R_{E}$ and $R_{\rm had}$ can be separately determined by fitting a sufficiently large sample of events covering a range of zenith angles. 

In this analysis we test the consistency of the observed and predicted ground signal \emph{event by event}, for a large sample of events covering a wide range of $X_{\rm max}$ and zenith angles.  By selecting simulated events which accurately match the observed LP of each event, we largely eliminate the noise from shower-to-shower fluctuations in the ground signal due to fluctuations in $X_{\rm max}$, while at the same time maximally exploiting the relative attenuation of the EM and muonic components of the shower.     

The LP and lateral distribution of the ground signal of an illustrative event are shown in Fig.~\ref{figFDSDComp}, along with a matching proton and iron simulated event;  the ground signal size is measured in units of vertical equivalent muons (VEM), the calibrated unit of SD signal size~\cite{augerSDVEMCal}.  Figure \ref{figFDSDComp} (bottom) illustrates a general feature of the comparison between observed and simulated events:  the ground signal of the simulated events is systematically smaller than the ground signal in the recorded events. 
Elucidating the nature of the discrepancy is the motivation for the present study.
 
The data we use for this study are the 411 hybrid events with $10^{18.8} < E < 10^{19.2}$ eV  and zenith angle $0^{\circ}-60^{\circ}$ recorded between 1 January 2004 and 31 December 2012, which satisfy the event quality selection criteria in Refs. \cite{sdfdCalICRC, sdfdCalICRC11}.  We thus concentrate on a relatively narrow energy range such that the mass composition changes rather little~\cite{augerXmaxMeas14,augerXmaxInterp14}, while having adequate statistics.  This energy range corresponds to an energy of 110 to 170 TeV in the center-of-mass reference frame of the UHECR and air nucleon, far above the LHC energy scale.   

\begin{figure}[t]
\centering
\includegraphics[width=\linewidth]{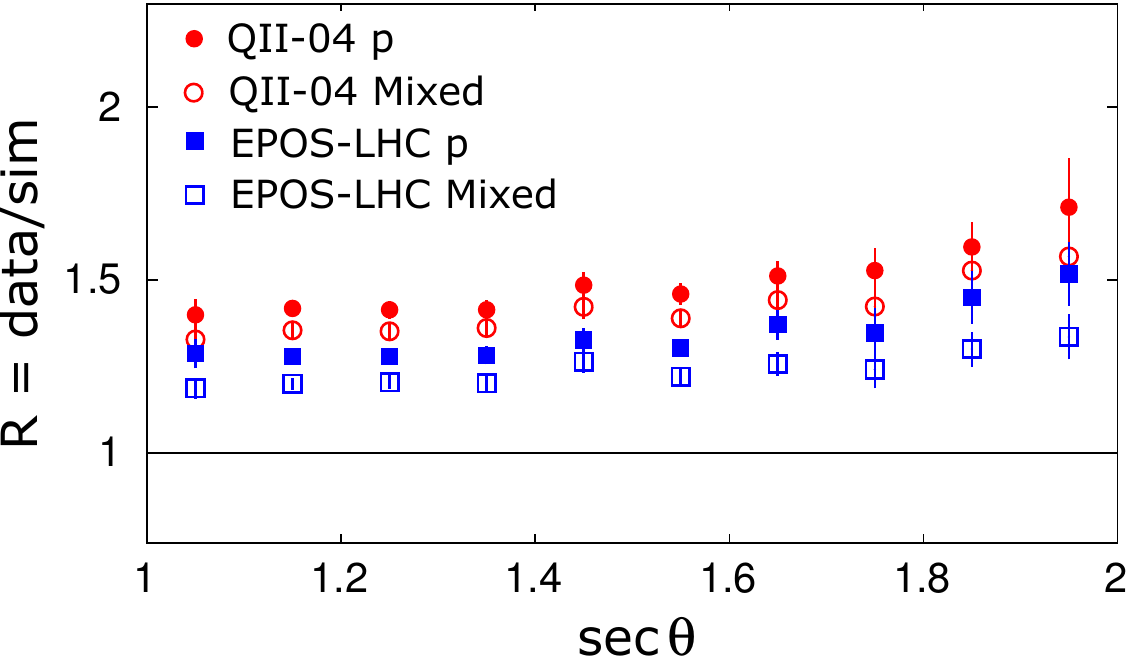}
\caption{The average ratio of S(1000) for observed and simulated events as a function of zenith angle, for mixed or pure proton compositions.}
\label{figSecTR}
\end{figure}

Figure \ref{figSecTR} shows the ratio of S(1000), the ground signal size at 1000 m from the shower core~\cite{augerNIM15}, for the events in our sample relative to that predicted for simulated events with matching zenith angle, depth-of-shower-maximum ($X_{\rm max}$) and calorimetric FD energy, for QGSJet-II-04~\cite{QII-04} and EPOS-LHC~\cite{EPOS-LHC}.  For each HEG, the analysis is done using the composition mix which reproduces the observed $X_{\rm max}$ distribution~\cite{augerXmaxMeas14,augerXmaxInterp14};  we also show the result for pure protons for comparison.    The discrepancy between a measured and simulated S(1000) evident in Fig.~\ref{figSecTR} is striking, at all angles and for both HEGs, and for both the mixed composition and pure proton cases.  

The zenith angle dependence of the discrepancy is the key to allowing $R_E$ and $R_{\rm had}$ 
to be separated.  As seen in Fig. \ref{components}, the ground signal from the hadronic component is roughly independent of zenith angle, whereas that of the EM component falls with sec($\theta)$, so that to reproduce the rise seen in Fig.~\ref{figSecTR}, the hadronic component must be increased with little or no modification of the EM component.  This will be quantified below.  

The analysis relies on there being no significant zenith-angle-dependent bias in the determination of the SD and FD signals.  The accuracy of the detector simulations 
as a function of zenith angle in the $0^{\circ}-60^{\circ}$ range of the study here, and hence the absence of a zenith angle dependent bias in the SD reconstruction, has been extensively validated with muon test data~\cite{tanktests}.  The absence of zenith angle dependence in the normalization of the FD signal follows from the zenith angle independence of $E_{\rm FD} /E_{\rm SD}$ of individual hybrid events.

\section{Production of Simulated Events}

The first step of the analysis is to generate a set of Monte Carlo (MC) events, to find simulated events matching the LPs of the data events.  The MC air-shower simulations are performed using the SENECA simulation code~\cite{Seneca}, with FLUKA~\cite{fluka1} as the low-energy HEG.  Simulation of the surface detector response is performed with GEANT4~\cite{geant4} within the software framework \Offline ~\cite{offline} of the Auger Observatory.   We produce showers matching each data event, with both HEGs and for all four primary cosmic-ray types (proton, helium, nitrogen, and iron nuclei), as follows: \\
$\bullet$ Repeatedly generate showers with the measured geometry and calorimetric energy of the given data event, reconstructing the LP and determining the $X_{\rm max}$ value until 12 showers having the same $X_{\rm max}$ value as the real event (within the reconstruction uncertainty) have been produced, or stopping after 600 tries.  For data events whose  $X_{\rm max}$ cannot be matched with all primary types, the analysis is done using only those primaries that give 12 events at this stage, in 600 tries \cite{Footnote1}.\\
$\bullet$ Repeat the simulation of these 12 showers at very high resolution, and select the 3 which best reproduce the observed longitudinal profile based on the  $\chi^2$-fit.  For each of the 3 selected showers, do a full surface detector simulation and generate SD signals for comparison with the data.  From these detailed simulations of 3 showers that match the full LP of the data event, determine the hadronic component of the simulated ground signal and the shower-to-shower variance.  

The choices of 12 and 3 showers in the two stages above assure, respectively, that i) the LPs of the final simulated data set fit the real data with a  $\chi^2$ distribution that is comparable to that found in a Gaisser-Hillas fit to the data itself, and ii) that the variance within the simulated events for a given shower is smaller than the shower-to-shower fluctuations in real events. More than $10^7$ showers must be simulated to create the analysis library of well-fitting simulated showers for the 411 hybrid events of the data set.  A high-quality fit to the LP is found for all events, for at least one primary type.

\section{Quantifying the Discrepancy}

\begin{figure}[t]
\centering
\includegraphics[width=\linewidth]{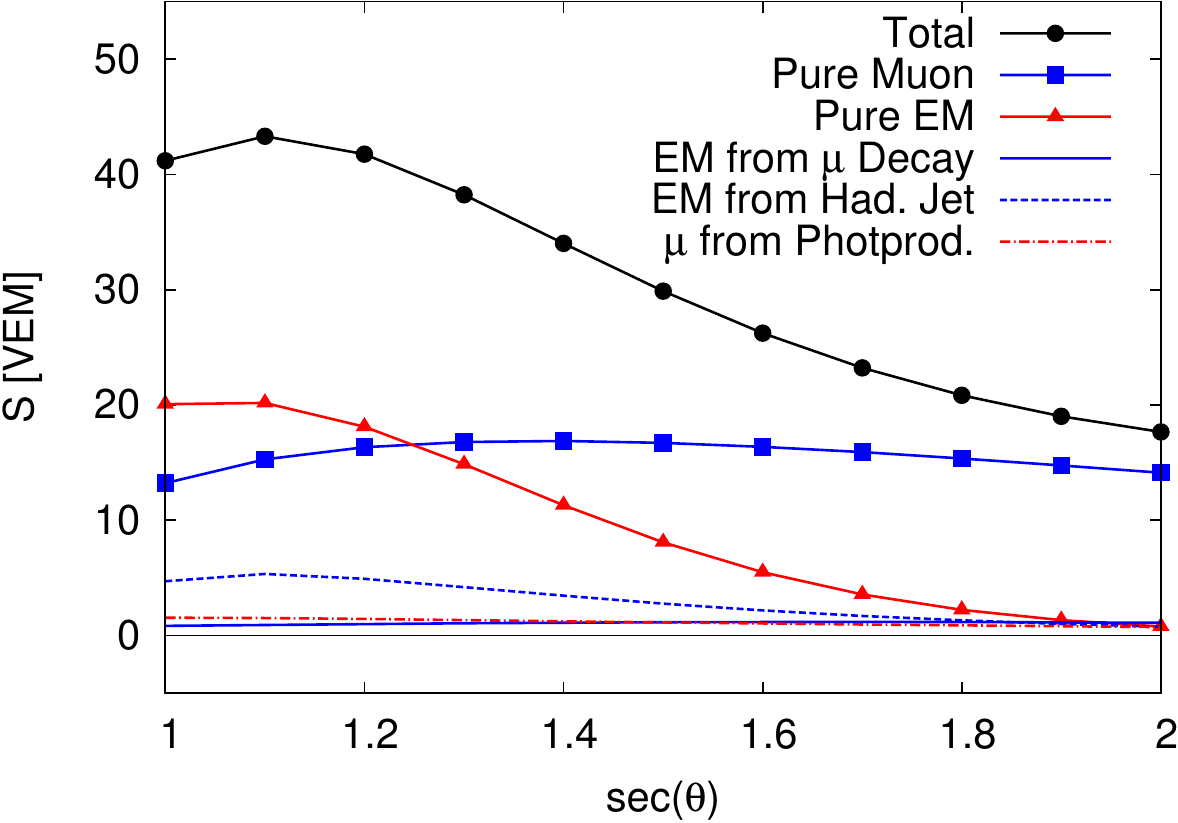}
\caption{The contributions of different components to the average signal as a function of zenith angle, for stations at 1 km from the shower core, in simulated 10 EeV proton air showers illustrated for QGSJet-II-04.}
\label{components}
\end{figure}

The history of all muons and EM particles ($e^{\pm}$ and $\gamma$'s) reaching the ground is tracked during simulation, following the description in Ref. \cite{univ2011}. Most muons come from $\pi^\pm$ or K decay and most EM particles from $\pi^0$ decay.  The portion of EM particles that are produced by muons through decay or radiative processes, and by low-energy $\pi^0$'s, are attributed to the hadronic signal, $S_{\rm had}$;  muons that are produced through photoproduction are attributed to the electromagnetic signal, $S_{EM}$.  The relative importance of the different components varies with zenith angle, as illustrated in Fig.~\ref{components}.   
Once $S_{EM}$ and $S_{\rm had}$ are known for a given shower $i$, with assumed primary mass $j$, the rescaled simulated S(1000) can be written as:
\begin{equation}
S_{\rm resc}(R_{E},R_{\rm had})_{i,j} \equiv R_{E} ~ {S_{EM, i,j}} +  R_{\rm had}  ~R_{E}^{\alpha} ~{S_{{\rm had}, i,j}}.
\label{Srs}
\end{equation}

The linear scaling of the EM contribution with $R_E$ is obvious, as is the factor $R_{\rm had}$ for the hadronic contribution.   The factor $R_{E}^{\alpha}$ reflects the fact that the hadronic signal increases slower than linearly with energy, since higher energy events require more stages in the shower cascade before the pions have low enough energy to decay to muons rather than re-interact, and at each stage, energy is removed from the hadronic cascade.  The value of $\alpha$ is a prediction of the HEG and depends also on mass; in practice both EPOS and QGSJet-II simulations find $\alpha \approx 0.9$, relatively independently of composition~\cite{alpha}.   We investigated the sensitivity of our conclusions to the possibility that $\alpha$ predicted by the models is incorrect, and find its potential effect is small enough to be ignored for the present analysis~\cite{Footnote2}.   

The best fit values of $R_{E}$ and $R_{\rm had}$ are determined by maximizing the likelihood function $\prod_{i} P_{i}$, where the index $i$ runs over each event in the data set and the contribution of the $i$th event is
\begin{equation}
\label{P_i}
P_{i} =  \sum_{j}   \frac{p_{j}\left(X_{{\rm max}, i}\right) }{\sqrt{2\pi}\sigma_{i,j}} ~ {\rm exp}\left[-\frac{\left(S_{\rm resc}(R_{E},R_{\rm had})_{i,j}-S(1000)_{i}\right)^2}{2~\sigma_{i,j}^2}\right].
\end{equation}
The index $j$ labels the different possible primaries (p, He, N and Fe), and $p_{j}\left(X_{{\rm max}, i}\right)$ is the prior on the probability that an event with $X_{{\rm max}, i}$ has mass $j$, given the mass fractions $f_j$ in the interval $10^{19\pm0.2}$ eV (see Ref. \cite{augerXmaxMeas14} for the fit to the observed $X_{\rm max}$ distribution for each HEG):  
\begin{equation}
p_j(X_{\rm max}) = f_j \, \mathcal{P}_j(X_{\rm max}) \,/  \,\Sigma_{j'} f_{j'} \, \mathcal{P}_{j'} (X_{\rm max}),
\end{equation}
where $\mathcal{P}_j(X_{\rm max})$ is the probability density of observing $X_{\rm max}$ for primary type $j$, for the given HEG.  
The variance entering Eq.~(\ref{P_i}) 
includes (a) measurement uncertainty of typically 12\%, from the uncertainty in the reconstruction of S(1000), the calorimetric energy measurement, 
and the uncertainty in the $X_{\rm max}$ scale, as well as (b) the variance in the ground signals of showers with matching LPs due to shower-to-shower fluctuations (ranging from typically $16$\% for proton-initiated showers to 5\% for iron-initiated showers)   
and (c) the uncertainty in separating $S_{\mu}$ and $S_{EM}$ in the simulation, and from the limited statistics of having only three simulated events (typically 10\% for proton-initiated showers and 4\% for iron-initated showers). \\

\section{Results and Discussion}

\begin{table}
\centering
\caption{$R_E$ and $R_{\rm had}$ with statistical and systematic uncertainties, for QGSJet-II-04 and EPOS-LHC.}
\label{tabFit}
\begin{tabular}{lcc}
  \hline\hline
  Model & $R_{E}$ & $R_{\rm had}$ \\
  \hline
QII-04 p & ~~$1.09 \pm 0.08 \pm 0.09$ & ~~$1.59 \pm 0.17 \pm 0.09$         \\
QII-04 Mixed & ~~$1.00 \pm 0.08 \pm 0.11$ & ~~$1.61 \pm 0.18 \pm 0.11$     \\
EPOS p &  ~~$1.04 \pm 0.08 \pm 0.08$ & ~~$1.45 \pm 0.16 \pm 0.08$      \\
EPOS Mixed  & ~~$1.00 \pm 0.07 \pm 0.08$ & ~~$1.33 \pm 0.13 \pm 0.09$  \\
    \hline\hline
\end{tabular}
\end{table}

\begin{figure}[t]
\centering
\includegraphics[width=\linewidth]{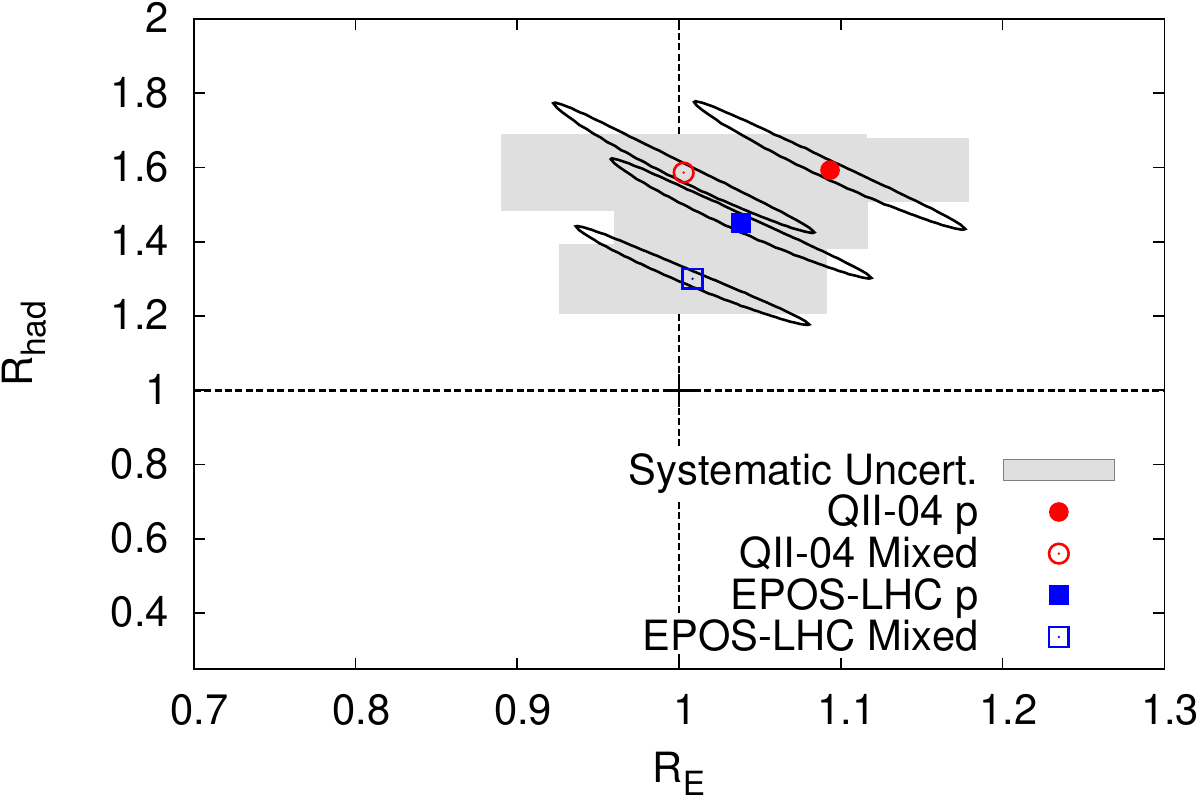}
\caption{Best-fit values of $R_{E}$ and $R_{\rm had}$ for QGSJet-II-04 and EPOS-LHC, for pure proton (solid circle/square) and mixed composition (open circle/square). The ellipses and gray boxes show the 1-$\sigma$ statistical and systematic uncertainties.}
\label{figContour}
\end{figure}

Table \ref{tabFit} gives the values of $R_{E}$ and $R_{\rm had}$ which maximize the likelihood of the observed ground signals, for the various combinations of HEGs and compositions considered.   The systematic uncertainties in the reconstruction of $X_{\rm max}$, $E_{\rm FD}$ and S(1000) are propagated through the analysis by shifting the reconstructed central values by their one-sigma systematic uncertainties.   Figure \ref{figContour} shows the one-sigma statistical uncertainty ellipses in the $R_{E}-R_{\rm had}$ plane; the outer boundaries of propagating the systematic errors are shown by the gray rectangles.   

The values of $R_{\rm had}$ needed in the models are comparable to the corresponding muon excess detected in highly inclined air showers~\cite{augerHorizMuons15}, as is expected because at high zenith angle the nonharonic contribution to the signal (shown with red curves in Fig.~\ref{components}) is much smaller than the hadronic contribution.  However the two analyses are not equivalent because a muon excess in an inclined air shower is indistinguishable from an energy rescaling, whereas in the present analysis the systematic uncertainty of the overall energy calibration enters only as a higher-order effect.  Thus the significance of the discrepancy between data and model prediction is now more compelling, growing from 1.38 (1.77) sigma  to 2.1 (2.9) sigma, respectively, for EPOS-LHC (QGSJet II-04), adding statistical and systematic errors from Fig. 6 of Ref. \cite{augerHorizMuons15} and Table \ref{tabFit}, in quadrature.

The signal deficit is smallest (the best-fit $R_{\rm had}$ is the closest to unity) with EPOS-LHC and mixed composition.  This is because, for a given mass, the muon signal is $\approx15 $\% larger for EPOS-LHC than QGSJet-II-04~\cite{pierogEPOSvsQII}, and in addition the mean primary mass is larger when the $X_{\rm max}$ data are interpreted with EPOS rather than with QGSJet-II~\cite{augerXmaxInterp14}.
  
Within the event ensemble used in this study, there is no evidence of a larger event-to-event variance in the ground signal for fixed $X_{\rm max}$ than predicted by the current models.  This means that the muon shortfall cannot be attributed to an exotic phenomenon producing a very large muon signal in only a fraction of events, such as could be the case if micro-black holes were being produced at a much-larger-than-expected rate~\cite{feng+BH01,faUHECR13}. 

\section{Summary}

We have introduced a new method to study hadronic interactions at ultrahigh energies, which minimizes reliance on the absolute energy determination and improves precision by exploiting the information in individual hybrid events.   We applied it to hybrid showers of the Pierre Auger Observatory with energies 6-16 EeV ($E_{\rm CM}$ = 110 to 170 TeV) and zenith angle $0^\circ-60^\circ$, to quantify the disparity between state-of-the-art hadronic interaction modeling and observed UHECR atmospheric air showers.  We considered the simplest possible characterization of the model discrepancies, namely an overall rescaling of the hadronic shower, $R_{\rm had}$, and we allow for a possible overall energy calibration rescaling, $R_{E}$.  

No energy rescaling is needed:  $R_{E} = 1.00\pm 0.10$ for the mixed composition fit with EPOS-LHC, and $R_{E} = 1.00\pm 0.14$ for QGSJet II-04, adding systematic and statistical errors in quadrature.  This uncertainty on $R_{E}$ is of the same order of magnitude as the 14\% systematic uncertainty of the energy calibration~\cite{sdfdCalICRC}. 

We find, however, that the observed hadronic signal in these UHECR air showers is significantly larger than predicted by models tuned to fit accelerator data.  
The best case, EPOS-LHC with mixed composition, requires a hadronic rescaling of $R_{\rm had} = 1.33\pm0.16$ (statistical and systematic uncertainties combined in quadrature), while for QGSJet II-04, $R_{\rm had} = 1.61\pm 0.21$.    
It is not yet known whether this discrepancy can be explained by some incorrectly modeled features of hadron collisions, possibly even at low energy, or may be indicative of the onset of some new phenomenon in hadronic interactions at ultrahigh energy.  Proposals of the first type include a higher level of production of baryons~\cite{pierogEPOSvsQII} or vector mesons~\cite{drescher07} (see Ref. \cite{engel15} for a recent review of the many constraints to be satisfied), while proposals for possible new physics are discussed in Refs. \cite{stringPerc12,faUHECR13,afICRC13}.  

The discrepancy between models and nature can be elucidated by extending the present analysis to the entire hybrid data set above $10^{18.5}$ eV, to determine the energy dependence of $R_E$ and $R_{\rm had}$.  In addition, the event by event analysis introduced here can be generalized to include other observables with complementary sensitivity to hadronic physics and composition, e.g., muon production depth~\cite{MPD}, risetime~\cite{risetime} and slope of the LDF.  

AugerPrime, the anticipated upgrade of the Pierre Auger Observatory~\cite{augerprime}, will significantly improve our ability to investigate hadronic interactions at ultrahigh energies, by separately measuring the muon and EM components of the ground signal.

\section*{Acknowledgments}

\begin{sloppypar}
The successful installation, commissioning, and operation of the Pierre Auger
Observatory would not have been possible without the strong commitment and
effort from the technical and administrative staff in Malarg\"ue.
\end{sloppypar}

\begin{sloppypar}
We are very grateful to the following agencies and organizations for financial
support:
\end{sloppypar}

\begin{sloppypar}
Comisi\'on Nacional de Energ\'\i{}a At\'omica, Agencia Nacional de Promoci\'on Cient\'\i{}fica
y Tecnol\'ogica (ANPCyT), Consejo Nacional de Investigaciones Cient\'\i{}ficas y
T\'ecnicas (CONICET), Gobierno de la Provincia de Mendoza, Municipalidad de
Malarg\"ue, NDM Holdings and Valle Las Le\~nas, in gratitude for their continuing
cooperation over land access, Argentina; the Australian Research Council;
Conselho Nacional de Desenvolvimento Cient\'\i{}fico e Tecnol\'ogico (CNPq),
Financiadora de Estudos e Projetos (FINEP), Funda\c{c}\~ao de Amparo \`a Pesquisa do
Estado de Rio de Janeiro (FAPERJ), S\~ao Paulo Research Foundation (FAPESP)
Grants No.\ 2010/07359-6 and No.\ 1999/05404-3, Minist\'erio de Ci\^encia e
Tecnologia (MCT), Brazil; Grant No.\ MSMT-CR LG13007, No.\ 7AMB14AR005, and the
Czech Science Foundation Grant No.\ 14-17501S, Czech Republic; Centre de Calcul
IN2P3/CNRS, Centre National de la Recherche Scientifique (CNRS), Conseil
R\'egional Ile-de-France, D\'epartement Physique Nucl\'eaire et Corpusculaire
(PNC-IN2P3/CNRS), D\'epartement Sciences de l'Univers (SDU-INSU/CNRS), Institut
Lagrange de Paris (ILP) Grant No.\ LABEX ANR-10-LABX-63, within the
Investissements d'Avenir Programme Grant No.\ ANR-11-IDEX-0004-02, France;
Bundesministerium f\"ur Bildung und Forschung (BMBF), Deutsche
Forschungsgemeinschaft (DFG), Finanzministerium Baden-W\"urttemberg, Helmholtz
Alliance for Astroparticle Physics (HAP), Helmholtz-Gemeinschaft Deutscher
Forschungszentren (HGF), Ministerium f\"ur Wissenschaft und Forschung, Nordrhein
Westfalen, Ministerium f\"ur Wissenschaft, Forschung und Kunst,
Baden-W\"urttemberg, Germany; Istituto Nazionale di Fisica Nucleare
(INFN),Istituto Nazionale di Astrofisica (INAF), Ministero dell'Istruzione,
dell'Universit\'a e della Ricerca (MIUR), Gran Sasso Center for Astroparticle
Physics (CFA), CETEMPS Center of Excellence, Ministero degli Affari Esteri
(MAE), Italy; Consejo Nacional de Ciencia y Tecnolog\'\i{}a (CONACYT) No.\ 167733,
Mexico; Universidad Nacional Aut\'onoma de M\'exico (UNAM), PAPIIT DGAPA-UNAM,
Mexico; Ministerie van Onderwijs, Cultuur en Wetenschap, Nederlandse
Organisatie voor Wetenschappelijk Onderzoek (NWO), Stichting voor Fundamenteel
Onderzoek der Materie (FOM), Netherlands; National Centre for Research and
Development, Grants No.\ ERA-NET-ASPERA/01/11 and No.\ ERA-NET-ASPERA/02/11,
National Science Centre, Grants No.\ 2013/08/M/ST9/00322, No.
2013/08/M/ST9/00728 and No.\ HARMONIA 5 -- 2013/10/M/ST9/00062, Poland;
Portuguese national funds and FEDER funds within Programa Operacional Factores
de Competitividade through Funda\c{c}\~ao para a Ci\^encia e a Tecnologia (COMPETE),
Portugal; Romanian Authority for Scientific Research ANCS, CNDI-UEFISCDI
partnership projects Grants No.\ 20/2012 and No.\ 194/2012, Grants No.
1/ASPERA2/2012 ERA-NET, No.\ PN-II-RU-PD-2011-3-0145-17 and No.
PN-II-RU-PD-2011-3-0062, the Minister of National Education, Programme Space
Technology and Advanced Research (STAR), Grant No.\ 83/2013, Romania; Slovenian
Research Agency, Slovenia; Comunidad de Madrid, FEDER funds, Ministerio de
Educaci\'on y Ciencia, Xunta de Galicia, European Community 7th Framework
Program, Grant No.\ FP7-PEOPLE-2012-IEF-328826, Spain; Science and Technology
Facilities Council, United Kingdom; Department of Energy, Contracts No.
DE-AC02-07CH11359, No.\ DE-FR02-04ER41300, No.\ DE-FG02-99ER41107 and No.
DE-SC0011689, National Science Foundation, Grants No.\ 0450696 and No.\ 1212528, and The Grainger
Foundation, USA; NAFOSTED, Vietnam; Marie Curie-IRSES/EPLANET, European
Particle Physics Latin American Network, European Union 7th Framework Program,
Grant No.\ PIRSES-2009-GA-246806; and UNESCO.
\end{sloppypar}


\begin{thebibliography}{99}
\bibitem{Footnote1} 261 of the 411 events have an $X_{\rm max}$ value so deep or shallow that the LP cannot be reproduced with all primaries for both HEGs, e.g., 209 (230) events cannot be described as Fe, and 32 (10) cannot be described as a proton, in EPOS-LHC (QGSJetII-04).
\bibitem{Footnote2} The potential sensitivity to $\alpha$ arises in spite of the narrow primary energy range, because individual nucleons in a primary of mass $A$ carry energy $E/A$. To investigate this we let $R_{\rm had}(A,E) \equiv (\frac{E}{10{\rm \,EeV}}/A)^{\delta \alpha} ~ R_{\rm had} $, where $R_{\rm had}$ is the rescaling factor for proton primaries at $10^{19}$\,eV. We take ${\delta \alpha}=0.01$ as the fiducial value, motivated by \cite{afICRC13} which achieves $R_{\rm had} \approx 1.3$ by converting $\pi^{0}$'s and $\eta$'s to other mesons, resulting in ${\delta \alpha} = 0.0135$ for EPOS and $ 0.01-0.02 $ for QGSJet.  Reanalyzing with ${\delta \alpha}=0 \, (0.02)$, $R_{E}$ changes negligibly and $R_{\rm had} \rightarrow 1.30 \,(1.37)$ for EPOS.
\end{thebibliography}
\end{document}